\documentclass[12pt]{article}
\usepackage{amsfonts}
\usepackage{amssymb}
\def\lp{\stackrel{\leftarrow}{\partial}}
\def\rp{\stackrel{\rightarrow}{\partial}}
\def\be{\begin{eqnarray}}
\def\ee{\end{eqnarray}}
\def\*{\star}
\begin{document}
          \begin{flushright}  
          ANL-HEP-PR-02-030 \\ MIAMI-TH-1-02
           \end{flushright}
{\LARGE
\centerline{Deformation Quantization of Superintegrable Systems}
\centerline{and Nambu Mechanics   } }

\phantom{aaa} 

\phantom{aaa} 

{\large
\centerline{Thomas L Curtright$^{\S}$  and Cosmas K Zachos$^{\P}$}  }
$^{\S}$ Department of Physics, University of Miami,
Box 248046, Coral Gables, Florida 33124, USA\\
\phantom{a} \qquad\qquad{\sl curtright@physics.miami.edu}  

$^{\P}$ High Energy Physics Division,
Argonne National Laboratory, Argonne, IL 60439-4815, USA \\
\phantom{a} \qquad\qquad{\sl zachos@hep.anl.gov} 

\begin{abstract} 
Phase Space is the framework best suited for quantizing superintegrable 
systems, naturally preserving the symmetry algebras of the 
respective hamiltonian invariants. The power and simplicity of the method 
is fully illustrated through new applications to nonlinear $\sigma$-models, 
specifically for de Sitter 
$N$-spheres and Chiral Models, where the symmetric quantum 
hamiltonians amount to compact and elegant expressions. 
Additional power and elegance is provided by the use of 
Nambu Brackets to incorporate the extra invariants of 
superintegrable models.  Some new classical results 
are given for these brackets, and their quantization is successfully
compared to that of Moyal, validating Nambu's original proposal. 

\end{abstract}
\noindent\rule{7in}{0.01in}

\section {Introduction}
Highly symmetric quantum systems are often integrable, and, in 
special cases, superintegrable and exactly solvable \cite{winter}.
A superintegrable system of $N$ degrees of freedom has more than
$N$ independent invariants, and a maximally superintegrable one has $2N-1$ 
invariants. 
In the case of velocity-dependent potentials, when quantization 
of a classical system presents operator ordering ambiguities 
involving $x$ and $p$, the general consensus has long been 
\cite{velo,lakshmanan,higgs,leemon} to select those orderings
in the quantum hamiltonian which maximally preserve the symmetries 
present in the corresponding classical hamiltonian. 
Often, even for simple systems, such as 
$\sigma$-models considered here, such constructions 
may become involved and needlessly technical.

It is pointed out here that, in contrast to conventional operator 
quantization, this problem of selecting the quantum hamiltonian which maximally
preserves integrability is addressed most suitably and cogently in Moyal's 
phase-space quantization formulation  \cite{moyal,cfz,czreview}. 
The reason is that the variables involved in it 
(``classical kernels" or ``Weyl transforms of operators") are c-number 
functions, like those of the classical phase-space theory, 
and have the same interpretation, although they involve $\hbar$-corrections
(``deformations"), in general---so $\hbar\rightarrow 0$ reduces them to the 
classical expressions.  It is only the detailed algebraic structure 
of their respective brackets and composition rules which contrast with the 
variables of the classical theory. This complete formulation is based on the 
Wigner Function (WF), which is a quasi-probability distribution function in 
phase-space, and comprises the kernel function of the density matrix.
Observables and transition amplitudes are  phase-space integrals of kernel
functions weighted by the WF, in analogy to statistical mechanics. Kernel
functions, however, unlike classical functions, compose through the
$\*$-product, a noncommutative, associative, pseudodifferential operation, 
which encodes the entire quantum mechanical action and whose antisymmetrization 
(commutator) is the Moyal Bracket (MB) \cite{moyal,cfz,czreview}. 

Any arbitrary operator ordering could be brought to Weyl-ordering format, by
use of Heisenberg commutations, and through Weyl's transform corresponds
invertibly to a specific $\hbar$-deformation in the classical kernel 
\cite{weyl,groenewold}. Thus,
two operators of different orderings correspond to kernel functions differing
in their deformation terms of $O(\hbar)$.  The problem thus reduces to a
purely $\*$-product algebraic one, as the resulting preferred orderings are
specified and encoded most simply by far through the particular deformation of
the resulting c-number kernel expressions.

Hietarinta \cite{hietarinta} has investigated in this phase-space quantization
language the simplest integrable systems of velocity-dependent potentials. In
each system, he has promoted the vanishing of the Poisson Bracket (PB) of the
(one) classical invariant $I$ (conserved integral)  with the hamiltonian, 
$\{ H,I\}=0$, to the vanishing of its (quantum) Moyal Bracket (MB) with the
hamiltonian, $\{\! \{ H_{qm} ,I_{qm} \}\!  \} =0$. 
This dictates quantum corrections, addressed
 perturbatively in $\hbar$: he has found $O(\hbar^2)$ corrections to the $I$s
and $H$ ($V$), needed for quantum symmetry. The expressions found are quite
simple, as the systems chosen are such that the polynomial character of the
$p$s, or suitable balanced combinations of $p$s and $q$s, ensure collapse or
subleading termination of the MBs. The specification of the symmetric
hamiltonian then is complete, since {\em the quantum hamiltonian in terms of
classical phase-space variables corresponds uniquely to the Weyl-ordered
expression for these variables in operator language}. Berry \cite{berry} has 
also studied the WFs of integrable systems in great depth.

In this paper, nonlinear $\sigma$-models (with explicit illustrations 
on $N$-spheres and chiral models)  are utilized to argue for the general 
principles of power and convenience in isometry-preserving quantization in
phase space, for large numbers of invariants, in principle (as many as the
isometries of the relevant manifold). In the cases illustrated, the number of
algebraically independent invariants matches or exceeds the dimension of the
manifold, leading to superintegrability \cite{winter}, whose impact is best 
surveyed through Nambu Brackets (NB) \cite{nambu,hietnambu,takhtajan,nutku}.
The procedure of determining the proper
symmetric quantum Hamiltonian then yields remarkably compact and elegant
expressions.

Briefly, we find that the symmetry generator invariants are undeformed by 
quantization, but the Casimir invariants  of their MB algebras are deformed.
Hence, the hamiltonians are also deformed by terms $O(\hbar^2)$, as they 
consist of quadratic Casimir invariants. Their spectra are then read off 
through group theory, properly adapted to phase space. The basic 
principles are illustrated for the simplest 
curved manifold, the two-sphere, in Section 2;
while generalization to larger classes of symmetric manifolds such as 
Chiral Models and $N$-spheres is provided in Sections 3 and 4, which also 
investigate the underlying distinctive geometry of such models. Moreover, 
in Section 5, the classical evolution of all functions in phase space 
for such systems is specified through NBs, whose quantization is briefly 
outlined, and compared to the the standard Moyal deformation quantization 
utilized in this work. This comparison validates Nambu's original 
quantization proposal. Conclusions are summarized in Section 7, while a few
geometrical derivations on the classical structure of chiral models are 
provided in the Appendix.

\section {Principles and $S^2$}
Consider a particle on a curved manifold, in integrable one-dimensional 
$\sigma$-models considered by R Sasaki (unpublished), 
\begin{equation}
L(q,\dot{q})={1\over2} g_{ab}(q)~\dot{q}^a\dot{q}^b,
\label{Posaunenstoss}
\end{equation}
so that 
\begin{equation}
p_a={\partial L\over{\partial \dot{q}^a}}= g_{ab}~ \dot{q}^b,
\qquad \qquad  \dot{q}^a= g^{ab} p_b~.
\end{equation}
Thus, 
\begin{equation}
H(p,q)={1\over2} g^{ab} p_a p_b  \quad   (=L).
\label{Amsel}
\end{equation}
The isometries of the manifold generate the conserved integrals of the 
motion \cite{bcz}. The classical equations of motion are 
\be
\dot{p}_a= -\frac{g^{bc}_{,a}}{2}~ p_b p_c = \frac{g_{bc,a}}{2}~ \dot{q}^b
 \dot{q}^c.
\ee

As the simplest possible nontrivial illustration, consider a particle on a 
2-sphere of unit radius, $S^2$.  In Cartesian coordinates (after the 
elimination of $z$, so $q^1=x, q^2=y$), one has, for $a,b=1,2$:
\begin{equation} 
g_{ab}= \delta_{ab}+ {q^a q^b \over u}, \qquad  \qquad  
g^{ab}= \delta_{ab}- q^a q^b , \qquad  \qquad   \det g_{ab} =\frac{1}{u},
\qquad   \qquad u\equiv 1-x^2-y^2~.
\end{equation}
($u$ is the sine-squared of the latitude, since this represents 
the orthogonal projection of the globe on its equatorial plane).
The momenta are then
\be
p_a= \dot{q}^a + q^a {h\over u}  
= \dot{q}^a + q^a (q\cdot p)  ~, \qquad   \qquad   
h\equiv -\dot{u}/2 = x \dot{x} + y \dot{y} ~. 
\ee
The classical equations of motion here amount to 
\be
\dot{p}_a = p_a ~q\cdot p ~, \qquad \hbox{i.e.,}  \qquad   
\ddot{q}^{~a}+ q^a \left ( {\dot{h}\over u} + {h^2 \over u^2}\right ) =0.
\ee

It is then easy to find the three classical invariants, the 
components of the conserved angular momentum in this nonlinear realization,
\be
L_z&=& x p_y - y p_x ~, \\
L_y&=& \sqrt {u} ~ p_x  ~, \nonumber \\
L_x&=&  -\sqrt {u}~ p_y ~.\nonumber 
\ee
(The last two are the de Sitter ``momenta", or nonlinearly realized 
``axial charges" corresponding to the ``pions" $x,y$ of the $\sigma$-model:
linear momenta are, of course, not conserved).
Their PBs close into $SO(3)$, 
\be
\{L_x,L_y \}= L_z ~,\qquad   \qquad   
 \{L_y,L_z \}= L_x ~,\qquad   \qquad    
\{L_z,L_x\}= L_y~. \label{Hacivat} 
\ee
Thus, it follows algebraically that their PBs with the Casimir invariant 
${\bf L}\cdot{\bf L}$ vanish. Naturally, since $H= {\bf L}\cdot{\bf L}/2$, 
they are manifested to be time-invariant, 
\be 
\dot{{\bf L}}=\{ {\bf L}, H \}=0.
\ee

In quantizing this system, operator ordering 
issues arise, given the effective velocity(momentum)-dependent potential.
In phase-space quantization, one may insert Groenewold's \cite{groenewold} 
associative (and non-commutative) $\*$-products, 
\be
\star \equiv \exp \left ( {i\hbar \over 2} 
(\lp_x \rp_{p_x} - \lp_{p_x} \rp_x +\lp_y \rp_{p_y} - \lp_{p_y} \rp_y ) 
\right ),
\ee
 in strategic points and orderings of the variables of (\ref{Amsel}), 
to maintain integrability. That is, the classical invariance 
expressions (PB commutativity),
\be
\{ I, H \}=0
\ee
are to be promoted to quantum invariances (MB commutativity),
\be
\{\!\{ I_{qm}, H_{qm}\}\!\} \equiv {I_{qm}\* H_{qm}- H_{qm}\* 
I_{qm}\over i\hbar} =0~.
\ee
(As $\hbar\rightarrow 0$ the MB reduces to the PB.)
Here, this argues for a (c-number kernel function) hamiltonian of the form
\be
H_{qm}= \frac{1}{2} \Bigl ( L_x\*  L_x +   
L_y\*  L_y +   L_z \*  L_z \Bigr ) .  \label{siornionios}
\ee
The reason is that, in this realization, the algebra (\ref{Hacivat}) is 
promoted to the corresponding MB expression {\em without any modification}, 
since all of its MBs collapse to PBs by the linearity in momenta of the 
arguments: all corrections $O(\hbar)$ vanish. 
Consequently, these particular invariants are undeformed by quantization, 
${\bf L}= {\bf L}_{qm}$. 
As a result, given associativity for $\*$, the corresponding quantum  
quadratic Casimir invariant
${\bf L}\cdot\* {\bf L}$ has vanishing MBs with ${\bf L}$ (but not vanishing 
PBs), and automatically serves as a symmetry-preserving hamiltonian.
The specification of the maximally symmetric quantum hamiltonian is thus 
complete.
 
The $\*$-product in this hamiltonian trivially evaluates to yield the quantum 
correction to (\ref{Amsel}),
\be
H_{qm} = H + \frac{\hbar^2}{8} \Bigl ( \det g -3 \Bigr ). \label{okanumaihi}
\ee

In phase-space quantization \cite{moyal,cfz,czreview}, the WF (the kernel 
function of the density 
matrix) evolves according to Moyal's equation \cite{moyal}, 
\be 
{\partial f \over \partial t} 
=\{\! \{H_{qm} , f \}\! \} ~; \label{evolution}
\ee
in addition to it, the WFs for pure stationary states also 
satisfy \cite{dbf,cfz} $\*$-genvalue equations specifying the spectrum,
\be 
H_{qm}  (x,p)\star f(x,p) &=& f(x,p)   \star H_{qm} (x,p)
 \nonumber \\
&=&H_{qm} \left (x+{i\hbar\over 2} \rp_p ~,~ 
p-{i\hbar\over 2} \rp_x\right ) ~ f(x,p)= E ~f(x,p)~. \label{stargenvalue}
\ee

The spectrum of this hamiltonian, then, is proportional to the $\hbar^2 l(l+1)$ 
spectrum of the $SO(3)$  Casimir invariant ${\bf L}\cdot\* {\bf L}=
L_{+}\* L_{-}+ L_z\* L_z - \hbar L_z$, for integer $l$ \cite{bffls}. It can 
be produced algebraically by the identical standard recursive ladder operations 
in $\*$ space which obtain in the operator formalism Fock space, 
\be
L_z  \* L_{+} - L_{+} \* L_z = \hbar L_{+}   ~, 
\ee
where $L_{\pm}\equiv L_x \pm  i L_y$. 

To bound the $\*$-spectrum of $L_z$, an adaptation of the standard argument is 
needed to expectation values which are WF-weighted phase-space
integrals in this formulation. Indeed, from the real $\*$-square theorem 
\cite{hbar}, it follows that
\be
\langle {\bf L}\cdot\* {\bf L}-  L_z\* L_z \rangle =
\langle L_x\* L_x +L_y\* L_y \rangle \geq 0.
\ee
The $\*$-genvalues of $L_z$, $m$, are thus bounded, 
$|m|\leq  l < \sqrt{ \langle {\bf L}\cdot\* {\bf L}\rangle}/\hbar  $, 
necessitating 
$L_{-} \* f_{m=-l} =0$.  Hence 
\be
 L_{+}\* L_{-}\* f_{-l}  =  0=(
{\bf L}\cdot\* {\bf L} - L_z\* L_z + \hbar L_z)\* f_{-l}  ~, 
\ee
and consequently $\langle {\bf L}\cdot\* {\bf L} \rangle=  \hbar^2 l(l+1)$. 
Similar $\*$-ladder arguments and inequalities apply directly in phase space 
to all Lie algebras.

Classical hamiltonians are scalar under canonical transformations, but it 
should not be assumed that the quantum mechanical expression 
(\ref{okanumaihi}) is a canonical scalar. If, instead of the orthogonal 
projection employed above, 
the gnomonic projection \cite{higgs} from the center of the sphere were used,
ie $PR_2$ projective coordinates, 
\be
Q^a= {q^a \over \sqrt{ 1-q^2} }~, \qquad   
Q^2+1 = {1 \over 1-q^2 }~, \qquad   
P^a= (p^a -q^a q\cdot p) \sqrt{1-q^2}~, 
\ee 
it would yield 
\be 
G^{ab}=\left(  1+Q^{2}\right)  \left(  \delta_{ab}+Q^{a}Q^{b}\right),
\qquad G_{ab} =\frac{1}{1+Q^{2}}\left(  \delta_{ab}
-\frac{Q^{a}Q^{b} }{1+Q^{2}} \right), \qquad 
\det G_{ab} =\frac{1}{\left(  1+Q^{2}\right)  ^3}~. 
\ee 
The Hamiltonian would now be polynomial,
\be
H=\frac{1}{2}\left(  1+Q^{2}\right)  
\left(P^2 +(Q\cdot P)^2 \right). 
\ee
Rewritten in terms of its invariants, 
\be
 L_Z= X P_Y - Y P_X ~, \qquad L_Y= P_X + X P\cdot Q  ~, \qquad
L_X=  -(P_Y + Y P\cdot Q) ~, 
\ee
which would obey the same MB $SO(3)$ algebra as before, 
it would specify a quantum hamiltonian
\be
H_{QM}\equiv \frac{1}{2}\left(  L_{X}\bigstar L_{X}+L_{Y}\bigstar L_{Y} 
+L_{Z}\bigstar L_{Z}\right),
\ee
where $\bigstar$ involves $Q,P$ instead of $q,p$. This then would lead to 
the polynomial quantum correction,  
\be
\frac{ \hbar^{2}}{8} \left(  5 Q^{2}  -2\right).  
\ee
But this would be different from the above correction, 
\be
\frac{ \hbar^{2}}{8} \left( {1\over 1-q^2} -3 \right). 
\ee
For canonical transformations in phase-space quantization see \cite{cfz}.
The $\*$-product and WFs would not be invariant, but would transform in a 
suitable quantum covariant way \cite{cfz}, so as to 
yield an identical MB algebra 
and $\*$-genvalue equations, and thus spectrum, following from the identical 
group theoretical construction. 

\section {Chiral Models} 
The treatment of the 3-sphere $S^3$ is very similar, with some
significant differences, since it also accords to the standard chiral model 
technology. The metric and eqns of motion, etc, 
are identical in form to those above, except now
$u\equiv 1-x^2-y^2-z^2= 1/\det g $,   
$~~h\equiv -\dot{u}/2 = x \dot{x} + y \dot{y}+z \dot{z}$,
and $a,b=1,2,3$. However, the description simplifies upon utilization
of Vielbeine, $g_{ab}= \delta_{ij} V_a^i V_b^j$ and 
$g^{ab} V_a^i V_b^j=\delta^{ij}$.

Specifically, the Dreibeine, are either left-invariant, or 
right invariant \cite{cuz}: 
\be
^{(\pm)}V_a^i=   \epsilon^{iab} q^b \pm  \sqrt{u}~ g_{ai} ~,
\qquad \qquad 
^{(\pm)}V^{ai}=   \epsilon^{iab} q^b \pm  \sqrt{u}~ \delta^{ai} .
\ee
The corresponding right and left conserved charges (left- and right-invariant,
respectively) then are
\be
R^i= ~^{(+)}V^i_a  ~ \dot{q}^a  = ~^{(+)}V^{ai} p_a ~, \qquad \qquad 
L^i=~^{(-)}V^i_a   ~ \dot{q}^a =~^{(-)}V^{ai} p_a ~.
\ee
More intuitive than those for $S^2$  are the linear combinations into Axial and 
Isospin charges (again linear in the momenta),
\be
{{\bf R}- {\bf L}\over 2}= \sqrt{u} ~ {\bf p}\equiv {\bf A} ,  \qquad
{{\bf R}+{\bf L}\over 2}= {\bf q}\times {\bf p}  \equiv {\bf I}.
\ee

It can be seen that the ${\bf L}$s and the ${\bf R}$s have 
PBs closing into standard $SU(2)\otimes SU(2)$, ie,  
$SU(2)$ relations within each set, and vanishing 
between the two sets. Thus they are seen to be constant, since the hamiltonian 
(and the lagrangian) can, in fact, be written in terms of either 
quadratic Casimir invariant,
\be
H= \frac{1}{2}  {\bf L}\cdot {\bf L}=
  \frac{1}{2}  {\bf R}\cdot {\bf R} =L  ~.     \label{specS3}
\ee

Quantization consistent with integrability thus proceeds as 
above for the 2-sphere, since the MB algebra collapses to PBs again,
and so the quantum invariants {\bf L} and {\bf R} again coincide with 
the classical ones, without deformation (quantum corrections). The 
$\*$-product is now the obvious 
generalization to 6-dimensional phase-space. The eigenvalues of the relevant
Casimir invariant are now $j(j+1)$, for half-integer $j$ \cite{chugoddard}.
However, this being a chiral model ($G\otimes G$), the symmetric
quantum hamiltonian is simpler that the previous one, since it can  now
also be written geometrically as 
\be
H_{qm}= \frac{1}{2} (p_a V^{ai}) \* ( V^{bi}  p_b) =
\frac{1}{2}\left (g^{ab} p_a p_b + {\hbar^2 \over 4} 
\partial_a V^{bi} \partial_b V^{ai} \right )  . \label{quantS3} 
\ee
The Dreibeine throughout this formula can be either $~^{+}V^i_a$ or
$~^{-}V^i_a$, corresponding to either the right, or the 
left-acting quadratic Casimir invariant.   
The quantum correction then amounts to 
\be
H_{qm}- H = {\hbar^2 \over 8} \Bigl ( \det g  - 7\Bigr ) .  \label{shozaburo}
\ee
This expression, ${\hbar^2 \over 8} (1/(1-q^2) - 7)$, 
again is not canonically invariant. Eg, in gnomonic $PR_3$ 
coordinates\footnote{The inverse gnomonic Vielbein is also polynomial, 
${\sf V}^{aj}=\delta^{aj}+Q^j Q^a +\epsilon^{jab} Q^b$.}, it is
$\frac{3}{4}\hbar^{2}(Q^{2}-1)$, ie, it has not transformed as a canonical 
scalar \cite{cfz}.

If one wished to interpret this simple result (\ref{quantS3}) in operator 
language 
(for operators ${\mathfrak x}$ and ${\mathfrak p}$), it would appear somewhat 
more complex: the first term,
$g^{ab}(x) p_a p_b/2$, would correspond to the Weyl-ordered expression
\be
\frac{1}{8}\Bigl ( {\mathfrak p}_a {\mathfrak p}_b g^{ab}({\mathfrak x}) 
+2 {\mathfrak p}_a  g^{ab}({\mathfrak x}) {\mathfrak p}_b + 
g^{ab}({\mathfrak x}) {\mathfrak p}_a {\mathfrak p}_b \Bigr )
=\frac{1}{2}  {\mathfrak p}_a  g^{ab} ({\mathfrak x}) {\mathfrak p}_b  + 
\frac{3\hbar^2 }{4} ~,    \label{opEckel}
\ee
in agreement with \cite{lakshmanan}. 
The second term, of course, is unambiguous, since it does not contain 
momenta.

\phantom{..} 

\phantom{..} 

In general, the above discussion also applies to all chiral models,
with $G\times G$ replacing $SU(2)\times SU(2)$ above. 
Ie, the Vielbein-momenta 
combinations $V^{aj}p_a$ represent algebra generator invariants, 
whose quadratic Casimir 
group invariants yield the respective hamiltonians, and whence the properly
$\*$-ordered quantum hamiltonians as above. (We follow the conventions of 
\cite{bcz}, taking the generators of $G$ in the defining representation to
be $T_j$.) 

That is to say, for 
\be
i U^{-1} \frac{d}{dt}   U=~^{(+)}V^j_a T_j \dot{q}^a =~^{(+)}V^{aj} p_a T_j  ~,
\qquad \qquad i U\frac{d}{dt}   U^{-1} =~^{(-)}V^{aj} p_a T_j ~,
 \ee
it follows that the PBs of the left- and right-invariant charges 
$~^{(\pm)}V^{aj} p_a =\frac{i}{2}Tr T_{j}U^{\mp 1}\frac{d}{dt}U^{\pm 1} $ 
close to the identical Lie algebras,
\be
\{ ^{(\pm)} V^{aj}p_a ,   ^{(\pm)}V^{bk}p_b \} = -2 f^{jkn}~^{(\pm)}V^{an}p_a ~,
\label{Lie}
\ee
and PB commute with each other, 
\be
\{ ~^{(+)} V^{aj}p_a ,  ^{(-)}V^{bk}p_b \}=0. \label{karagoz} 
\ee
These two statements are proved in the Appendix.

MBs collapse to PBs by linearity in momenta as before, and the hamiltonian is 
identical in form to (\ref{quantS3}). From (\ref{BCZ3.16}) of the Appendix,
 the quantum correction in  (\ref{quantS3}) is seen to amount to 
\be
H_{qm}- H = {\hbar^2 \over 8}\left (\Gamma^b_{ac}~ g^{cd} \Gamma^a_{bd} -
f_{ijk}f_{ijk}\right ) ,  \label{Plechazunga}
\ee
(reducing to (\ref{shozaburo}) for $S^3$). 

In operator language, this hamiltonian $H_{qm}$ amounts to 
Weyl-ordering of all products on the r.h.s., but, for generic groups, 
the first term in (\ref{quantS3}) does not reduce as simply as 
in (\ref{opEckel}) above. The spectra are given by 
the Casimir eigenvalues for the relevant 
algebras and representations.

\section {$S^N$} 
For the generic sphere models, $S^N$, the maximally symmetric hamiltonians 
are the quadratic Casimir invariants of $SO(N+1)$, 
\be
H=\frac{1}{2}  P_a P_a + \frac{1}{4} L_{ab} L_{ab} ~,  
\ee
where 
\be
P_{a}=\sqrt{u} ~p_a~, \qquad \qquad L_{ab}=q^a p_{b}-q^{b} p_a~, 
\ee
for $a=1,\cdots,N$, 
the de Sitter momenta and angular momenta of $SO(N+1)/SO(N)$.
All of these $N (N+1)/2$ sphere translations and rotations are symmetries of 
the classical hamiltonian. 
  
Quantization proceeds as in $S^2$,  maintaining conservation of all $P_{a}$ and
$L_{ab}$,
\be
H_{qm}=\frac{1}{2}  P_a \* P_a + \frac{1}{4} L_{ab} \* L_{ab} ~, 
\label{Moosbrugger}
\ee
and hence the quantum correction is 
\be
H_{qm}- H = {\hbar^2 \over 8}\left ( \frac{1}{u}- 1 -N(N-1) \right) .\label{sN}
\ee
The spectra are proportional to the 
Casimir eigenvalues $l(l+N-1)$ for integer $l$ \cite{bffls}. For $N=3$ 
of the previous section, 
this form is reconciled with the Casimir expression for (\ref{specS3}) as
$l=2j$, and agrees with \cite{lakshmanan,higgs,leemon}.

A plausible question might arise at this point, whether 
the above quantum hamiltonian (\ref{Moosbrugger}) could be expressed 
geometrically, in
tangent-space, as was detailed for the chiral models in the previous section.
For the generic sphere models, $S^N$,  the Vielbeine are  
\be
V_a^i= \delta_{ai} -\frac{q^a q^i}{q^2}\left ( 1\pm\frac{1}{\sqrt{u}}\right ),
\qquad V^{ai}=\delta_{ai} -\frac{q^a q^i}{q^2}\left( 1\pm \sqrt{u} \right ).
\label{viel}
\ee
The classical hamiltonian also equals 
\be
H=\frac{1}{2} (p_a V^{ai}) ( V^{bi}  p_b),
\ee
but the quantum hamiltonian (\ref{Moosbrugger}) is not equal to the chiral model
form,  
\be
H_{qm}\neq H'_{qm}\equiv \frac{1}{2} (p_a V^{ai}) \* ( V^{bi}  p_b).
\ee
Equality even fails for the $S^3$ case 
of the previous section, as (\ref{quantS3}) only holds for Dreibeine defined 
differently, as in that section. 

The cotangent bundle currents, for a general manifold, 
do not have their MBs reduce down to the Vielbein-currents as in (\ref{Lie}),
but, instead,
\be 
\left\{\!  \left\{  V^{aj}p_{a},V^{bk}p_{b}\right\}\!  \right\} 
=\omega^{a[jk]}p_{a}=\left(  V^{bk}\partial_{b}V^{aj}-V^{bj}\partial_{b}%
V^{ak}\right)  p_{a}  ~, 
\ee 
where, for the $N$-sphere, choosing the $-$ sign in (\ref{viel}), 
so $V^{aj}=\delta_{aj} - q^a q^j w$ for $w\equiv (1-\sqrt{1-q^{2}} )/q^2$, 
\be
\omega^{a[ij]}=w \left(\delta^{ai}q^{j} -  \delta^{aj}q^{i} \right)
~.
\ee
It follows that 
\be
H_{qm}-H'_{qm}=\frac{\hbar^{2} }{8}\left(  N-1\right)  \left( 1 -2w-N\right)  
=\frac{\hbar^{2}}{8}\left(  N-1\right)  \left(1+
\frac{2}{q^{2}}\left( \sqrt{1-q^{2}}  -1\right)  -N\right).
\label{Difference}%
\ee
$H'_{qm}$ corresponds to a different operator ordering 
in the conventional Hilbert space formulation, and has less symmetry 
than $H_{qm}$. $H'_{qm}$ conserves the rotation generators 
$L_{ab}$ (ie, it is symmetric under the 
$SO( N)$ stability subgroup for $S^{N}$); however, it does not
conserve the Vielbein-currents on the N-sphere, nor does it 
conserve the de Sitter momenta. This last statement
follows from the difference $H_{qm}-H'_{qm}$ in (\ref{Difference}) dictating,
\be 
\left\{\!\left\{  
H_{qm}-H'_{qm},P_{c}\right \}\!\right\}  =\hbar^{2} q^c (N-1) \left(
\frac{1-2w}{4q^2}\right) ,
\ee
ie, $\{\!\{  H'_{qm},P_{c}\}\!\} \neq 0$.
Still, even with reduced symmetry, $H'_{qm}$ is maximally superintegrable for
$N\geq 5$.

Nevertheless, despite these differences, it can be shown 
that a $\*$-similarity transformation bridges 
these hamiltonians. Consider 
\be
w^{-\frac{(N-1)}{2}} \star p_{a}V^{aj}\star w^{\frac{(N-1)}{2}} 
=w^{-(N-1)}\star\left(
p_{a}V^{aj} w^{N-1} \right)  
=V^{aj}p_{a}-i\hbar \frac{N-1}{2}V^{aj}\partial_{a}\ln w,
\label{punch}
\ee
and the complex conjugate  transformation 
\be
w^{\frac{(N-1)}{2}} \star p_{a}V^{aj}\star w^{-\frac{(N-1)}{2}} 
=\left(w^{N-1}  V^{aj}p_{a}\right)
\star w^{-(N-1)}=V^{aj}p_{a}+i\hbar \frac{N-1}{2}V^{aj}\partial_{a}\ln w.
\label{judy}
\ee
Associativity of the $\*$-product then allows the maximally symmetric real 
hamiltonian to be written as one half of
(\ref{judy})$\*$(\ref{punch}),
\be
H_{qm}=\frac{1}{2}\left(  w^{N-1}V^{aj}p_{a}\right)  \star 
w^{-2 (N-1) }  \star\left(  w^{N-1}V^{bj}p_{b}\right).
\ee
This form was discovered by using homogeneous 
coordinates on the sphere, with $1/w =1+\cos (\theta)$, 
where $\theta$ is the polar angle.

\section {Maximal Superintegrability and the Nambu Bracket} 

All the models considered above have extra invariants 
beyond the number of conserved quantities in involution
(mutually commuting) required for integrability in the Liouville sense. 
The most systematic way of accounting for such additional invariants, 
and placing them all on a more equal footing, even when they do not all 
simultaneously commute, is the NB formalism.  
 
For example, the classical mechanics of a particle on an N-sphere
as discussed above may be summarized elegantly through Nambu Mechanics in 
phase space \cite{nambu,takhtajan}. Specifically, \cite{hietnambu,nutku},
 in an $N$-dimensional space, and thus $2N$-dimensional 
phase space, motion is confined on the constant surfaces specified by 
the algebraically independent integrals of the motion (eg, $L_x,L_y,L_z$ 
for $S^2$ above.) Consequently, the phase-space velocity 
${\bf v}=(\dot{{\bf q}},\dot{{\bf p}})$  is always perpendicular to the 
$2N$-dimensional phase-space gradients $\nabla=(\partial_{\bf q}, 
\partial_{\bf p})$ of all these integrals of the motion. 

As a consequence, if there are $2N-1$ algebraically independent
such integrals, possibly including the hamiltonian, 
(ie, the system is maximally superintegrable \cite{winter}), 
the phase-space velocity
must be proportional \cite{hietnambu} to the cross-product of all those 
gradients, and hence the motion is fully specified for any phase-space 
function $k({\bf q},{\bf p})$
 by a phase-space Jacobian which amounts to the Nambu Bracket:
\be 
{dk\over dt}&=&\nabla k \cdot {\bf v}  \label{muhacir} \\
&\propto &\partial_{i_1} k ~ \epsilon^{i_1i_2...i_{2N}} ~ 
\partial_{i_2} L_{i_1} ... \partial_{i_{2N}}L_{2N-1} \nonumber   \\
& = & {\partial (k,L_1,...,...,L_{2N-1}) \over \partial 
(q_1,p_{1},  q_2,p_{2},  ...,q_{N},p_{N})   } \nonumber    \\
&\equiv & \{k,L_1,...,L_{2N-1}\}.\nonumber   
\ee 

For instance, for the above $S^2$, 
\be
{dk\over dt}= {\partial (k,L_x,L_y,L_z ) \over \partial 
(x,p_x,y,p_y)   } ~.  \label{Rum}
\ee
For the more general $S^N$, one now has a choice of $2N-1$ of the 
$N(N+1)/2$ invariants of $SO(N+1)$; one of several possible expressions is 
\be 
\frac{dk}{dt}= { \left(  -1\right)  ^{\left(  N^{2}-1\right)}  \over 
 P_{2}P_{3}\cdots P_{N-1} }
\frac{\partial\left(  k,P_{1},L_{12},P_{2},L_{23},P_{3},\cdots,P_{N-1}, 
L_{N-1\;N},P_{N}\right)  }{\partial\left(  x_{1},p_{1},x_{2},p_{2},
\cdots,x_{N},p_{N}\right)  } ~, \label{omunene}
\ee
where $P_{a}=\sqrt{u} ~ p_a$, for $a=1,\cdots,N$, 
 and $L_{a,a+1}=q^a p_{a+1}-q^{a+1} p_a$, for $~a=1,\cdots,N-1$. 

In general \cite{takhtajan}, NBs, being Jacobian determinants, 
possess all antisymmetries of such; being linear in all derivatives, they 
also obey the Leibniz rule of partial differentiation, 
\be
\{ k(L,M) ,f_1,f_2,... \} = 
\frac{\partial k}{\partial L} \{ L ,f_1,f_2,... \}   +
\frac{\partial k}{\partial M} \{ M ,f_1,f_2,... \} . \label{Leibniz}
\ 
\ee
Thus, an entry in the NB algebraically dependent on  the remaining entries 
leads to a 
vanishing bracket.  For example, it is seen directly from above that
the hamiltonian is constant, 
\be
{dH\over dt}=\left\{ { {\bf L}\cdot{\bf L} \over 2}, ...\right \}=0,
\ee
since each term of this NB vanishes. Naturally, this also applies to
all explicit examples discussed here, as they are all maximally 
superintegrable.

Finally, the impossibility to antisymmetrize 
more than $2N$ indices in $2N$-dimensional phase space, 
\be
\epsilon ^{ab....c [ i } \epsilon^{j_1 j_2 ...j_{2N}  ]   } =0,
\ee
leads to the Fundamental Identity, \cite{takhtajan},  
slightly generalized here, 
\be
\{ f_0 \{f_1,...,f_{m-1},f_m \} , f_{m+1},...,f_{2m-1}   \} +
\{f_m, f_0\{f_1,...,f_{m-1},f_{m+1}\},f_{m+2},...,f_{2m-1}\}\label{FI}  \\ 
+...+ \{f_m,...,f_{2m-2}, f_0 \{f_1,...,f_{m-1},f_{2m-1} \} \} =
\{ f_1,...,f_{m-1},f_0 \{ f_m , f_{m+1},...,f_{2m-1} \}   \}. \nonumber 
\ee
This $m+1$-term identity works for any $m$, and not just $m=2N$ here. 

The proportionality constant $V$ in (\ref{muhacir}),
\be
{dk\over dt}= V \{k,L_1,...,L_{2N-1} \}, \label{tuzsuz} 
\ee
has to be a time-invariant \cite{nutku} if it has no explicit time 
dependence. This is seen from consistency of (\ref{tuzsuz}),
application of which to 
\be
{d \over dt} (V\{f_1,...,f_{2N} \})= \dot{V}\{f_1,...,f_{2N}  \} +
V\{\dot{f_1},...,f_{2N} \} +...+
V\{f_1,...,\dot{ f}_{2N} \},
\ee
yields  
\be
V\{V\{f_1,...,f_{2N} \},L_1,...,L_{2N-1}\} &=& \dot{V}\{f_1,...,f_{2N} \} \\
+V\{V \{f_1,L_1,...,L_{2N-1} \} ,...,f_{2N} \} &+&...+
V\{f_1,..., V \{f_{2N} ,L_1,...,L_{2N-1} \}\}, \nonumber
\ee
and, by virtue of (\ref{FI}), $\dot{V}=0$ follows.

Closure under PBs of quantities serving as arguments in the NB
does {\em not} suffice for a NB to vanish, as illustrated in (\ref{Rum})
where $\{L_x,L_y \}= L_z$.
On the other hand, it is always true that PBs of 
conserved integrals are themselves conserved integrals, i.e., 
\be
{d\{L_a,L_b \}\over dt} \propto  \{ \{L_a,L_b\},L_1,...,L_{2N-1} \}  
\ee
must vanish. 

Actually, PBs result from a maximal reduction of NBs, 
by inserting $2N-2$ phase-space coordinates and summing over them, thereby
taking {\em symplectic traces}, 
\be
\left\{  L,M\right\}  =\frac{1}{\left(  N-1\right)  !}\left\{  L,M,x_{i_{1}},
p_{i_{1}},\cdots,x_{i_{N-1}},p_{i_{N-1}}\right\},
\ee 
where summation over all $N-1$ pairs of repeated indices is 
understood\footnote {Thus, from the identity (\ref{FI}), it follows that
\[
\{ \{L_a,L_b\},L_1,...,L_{2N-1} \}+
\{ L_b, \{L_a,L_1\},L_2,...,L_{2N-1} \}+...+
\{L_b, L_1,..., \{L_a,L_{2N-1} \}\}= \{L_a,\{ L_b, L_1,...,L_{2N-1} \}\} .
\]
Each NB in this consistency relation vanishes separately.}.
Fewer traces lead to relations between NBs of maximal rank, $2N$, and those
of lesser rank, $2k$,
\be
\left\{  L_{1},\cdots,L_{2k}\right\}  =\frac{1}{\left(  N-k\right)  !}\left\{
L_{1},\cdots,L_{2k},x_{i_{1}},p_{i_{1}},\cdots,x_{i_{N-k}},p_{i_{N-k}%
}\right\}
\ee 
(which is one way to define the lower rank NBs for $k\neq1$), or
between two lesser rank NBs. A complete theory of these relations has not
been developed; but, essentially,  $\left\{  L_{1},\cdots,L_{2k}\right\}$ acts
like a Dirac Bracket (DB) up to a normalization, $\{  L_{1} ,L_{2}\}  _{DB}$, 
where the fixed additional entries $L_{3},\cdots,L_{2k}$ in the
NB play the role of the constraints in the DB. (In effect, this has been
previously observed, e.g., \cite{takhtajan,nutku}, for the extreme case $N=k$,
without symplectic traces.)

As a simplest illustration, consider $N=k=2$ for the system (\ref{Rum}),
but now taking $L_x,L_y$ as second-class constraints:
\be
\{f,g,L_x,L_y\}= \{f,g\}  \{L_x,L_y \}+\{f,L_x\}  \{L_y,g\}- 
\{f,L_y\}  \{L_x,g \}\equiv  \{ L_x ,L_y\} ~\{f,g\}_{DB}.
\ee
That is, 
\be
\{f,g\}_{DB}= \Bigl (\{L_x,L_y\}\Bigr ) ^{-1}~   \{f,g,L_x,L_y\},  
\ee
so that, from (\ref{FI}) with $f_0= (\{L_x,L_y\})^{-1}=1/L_z$ (also see 
\cite{nutku}), it follows that the 
Dirac Brackets satisfy the Jacobi identity,
\be
\{ \{f,g\}_{DB},h\}_{DB}+
\{ \{g,h\}_{DB},f\}_{DB}+
\{ \{h,f\}_{DB},g\}_{DB}=0~,
\ee
a property usually established by explicit calculation \cite{bracketpamd}, in 
contrast to this derivation. Naturally, \linebreak 
$\{f,L_x,L_y,L_z\}=\{f,H \}_{DB}$.

By virtue of this symplectic trace, for a general system---not only 
a superintegrable one---Hamilton's 
equations admit an NB expression different than (\ref{tuzsuz}), 
\be
{dk\over dt}= \{ k,H \} = {1\over (N-1)!} 
 \{k,H,x_{i_1}, p_{i_1}, ...,x_{i_{N-1}},p_{i_{N-1}}\}. \label{papingo}
\ee

Despite considerable progress in the last six years \cite{dfst}, 
the deformation quantization of the Nambu formalism is not completely settled:
a transparent, user-friendly technique is not at hand. 
One desirable feature would be a quantized NB which reduces to MBs through 
symplectic traces. One might call such a deformation {\em autologous}
to the $\*$-product method applied throughout this paper.
This is not the case for the abelian deformation \cite{dfst}. But it is the
case for another, older approach to quantization, namely the method
considered by Nambu in an operator context \cite{nambu}, when applied 
to the phase-space formalism. 

Define Quantum Nambu Brackets (QNB),
\be 
\left [ A, B \right ]_{\*}   &\equiv &  i\hbar~ \{\!\{A,B \}\!\} \nonumber \\
\left [ A,B,C \right ] _{\*} &\equiv & A \star B\* C-A \* C \star B
+B\star C\star A-B\star A\star C+C\star A\star B-C\star B\star A \nonumber \\
\left[ A,B,C,D\right] _{\star } &\equiv &A\star \left[ B,C,D\right]
_{\star }-B\star \left[ C,D,A\right] _{\star }+C\star \left[
D,A,B\right] _{\star }-D\star \left[ A,B,C\right] _{\star }  \nonumber \\
&=&  [A,B]_\star \star [C,D]_\star + [A,C]_\star \star [D,B]_\star 
+ [A,D]_\star \star [B,C]_\star \nonumber \\ 
&\phantom{a} & +[C,D]_\star \star [A,B]_\star + [D,B]_\star \star [A,C]_\star 
+ [B,C]_\star \star [A,D]_\star~,   \label{QNB} 
\ee 
etc., and use these symmetrized $\*$-products in the quantum theory instead
of the previous jacobians. 

This approach grants only one of the three
mathematical desiderata: full antisymmetry. The Leibniz property 
(\ref{Leibniz}) and the Fundamental Identity (\ref{FI}) are not satisfied,
in general. To some extent, the loss of the latter two properties is a 
subjective shortcoming, and dependent on the specific application context. 
But, objectively, this approach is in
agreement with the $\*$-product quantization of the examples given above.

For example, by virtue of the MBs for the two-sphere, expressed in gnomonic
coordinates, 
\be 
\frac{dX}{dt}  & =&\left\{  \left\{  X,H_{QM}\right\}  \right\}  =\left(
1+Q^{2}\right)  \left(  P_{X}+\left(  P\cdot Q\right)  X\right) \nonumber \\
\frac{d P_X}{dt}& =&\left\{  \left\{  P_{X},H_{QM}\right\}  \right\}
  =-P_{X}\left(
P\cdot Q\right)  \left(  1+Q^{2}\right)  -X\left(  P^{2}+\left(  P\cdot
Q\right)  ^{2}\right)  -\frac{5}{4}\hbar^{2}X ~ .
\ee 
The first of these is classical in form, while the second
contains a quantum correction that is a hallmark of the method. 
By comparison, we also find exactly
the same results using Nambu's approach%
\be 
\frac{dX}{dt}=\frac{-1}{2\hbar ^{2}}\left[ X,L_{X},L_{Y},L_{Z}\right]
_{\bigstar }\;,\qquad \qquad \frac{dP_{X}}{dt}=\frac{-1}{2\hbar ^{2}}\left[
P_{X},L_{X},L_{Y},L_{Z}\right] _{\bigstar }\;,
\ee 
where the second of these includes the quantum correction $-5\hbar ^{2}X/4$
as above. 

In fact, 
this result generalizes to arbitrary functions of phase space 
with no explicit time dependence, for all coordinate frames.
Specifically, for $S^2$, it follows directly from (\ref{QNB}) and 
(\ref{Hacivat})  (with MBs supplanting PBs, $\{\!\{ L_x,L_y \}\! \} = L_z$,
$\{\!\{  L_y,L_z \}\! \}= L_x $,
$\{\!\{ L_z,L_x\}\! \}= L_y$) that the Moyal Bracket with the hamiltonian 
(\ref{siornionios}) equals Nambu's QNB, for an arbitrary function $k$ of phase 
space,
\be 
\left[ k,L_x ,L_y ,L_z \right] _{\*} 
=  i \hbar ~ \left[ ~k, {\bf L}\cdot\* {\bf L} ~\right]_{\star } 
= -2 \hbar^2 ~  \{\! \{ k, H_{qm} \}\!\} ~,
\ee 
so that, 
\be
\frac{dk}{dt}   =\frac{-1}{2\hbar ^{2}}\Bigl [ k,L_{x},L_{y},L_{z}\Bigl ]_{\*}.
\ee
For $\hbar\rightarrow 0$, it naturally goes to (\ref{Rum}).

As a derivation, this ensures that consistency
requirements (\ref{Leibniz}) and (\ref{FI}) {\em are} satisfied, with the 
suitable insertion of $\*$-multiplication in the proper locations to ensure
full combinatoric analogy,
\be
[ A\* B, L_x , L_y , L_z  ]_\*= A\* [ B, L_x , L_y , L_z  ]_\*  
+ [ A, L_x , L_y , L_z  ]_\* \* B~,  \label{qLeibnz}
\ee
and  
$$  
\!\!\!\!
[ [ L_x ,L_y ,L_z  ,D ]_\* ,E,F,G ]_\*  +  
[ D, [ L_x ,L_y ,L_z  ,E ]_\* ,F,G ]_\* +  
[ D,E, [ L_x ,L_y ,L_z  ,F ]_\* ,G ]_\* +  
[ D,E,F, [ L_x ,L_y ,L_z  ,G ]_\* ]_\*
$$
\be
  =  
[ L_x ,L_y ,L_z  , [ D,E,F,G ]_\* ]_\*  ~.
\ee

The reader might also wish to note from (\ref{QNB}) that, for any phase-space 
constant $A$, 
\be 
[ A,B,C,D ]_{\*}=0 
\ee
holds identically, in contrast to the 3-argument QNB \cite{nambu}.
Thus, $dA/dt=0$ is consistent, and {\em no} debilitating constraint 
among the arguments $B,C,D$ is imposed;  the inconsistency identified 
in ref \cite{nambu} is a feature of
odd-argument QNBs and does not restrict the even-argument 
QNBs of phase space considered here.

By contrast, one might try to define a quantized Nambu bracket $\left\{
\left\{ ,,,\right\} \right\} _{\bigstar }$ simply by taking $\*$-products of
the phase-space gradients that appear in the classical NB and applying Jordan's
trick of symmetrizing all such products at the expense of making the algebra
non-associative. This also fails to grant all three mathematical 
desiderata (antisymmetry, Leibniz property, and FI). But, more
importantly, it does not give the same equations of motion. Although,
in gnomonic coordinates,  
$dX/dt$ is as given above, the other equation of motion would now become 
\be
\frac{dP_X}{dt}=\left\{ \left\{ P_{X},L_{X},L_{Y},L_{Z}\right\} \right\}
_{\bigstar }=\left\{ \left\{ P_{X},H_{QM}\right\} \right\} _{\bigstar
}+\frac{3}{2}\hbar^{2} X ~. 
\ee 
Thus, in general, quantum corrections differ in these various methods.

In practice, however, given the simple energy spectrum and other features of 
the usual Moyal $\star$-product quantization (essentially, its equivalence to 
standard Hilbert-space quantum mechanics), it is clearly 
the preferred method for conventional problems such as the ones solved 
in this paper. In any case, quantum deformations of the NB should not only 
link (\ref{evolution}) up with (\ref{papingo}) and (\ref{tuzsuz}), as above,
but also provide equivalents to the $\star$-genvalue eqn (\ref{stargenvalue}) 
for static WFs, needed to support the spectral theory in such a formalism.

As indicated, in general, the QNB (which provide the 
correct quantization rule for the systems considered)  need not satisfy the 
Leibniz property and FI for consistency, as they are not necessarily 
derivations.  E.g., for $S^3$, to quantize (\ref{omunene}) 
for $N=3$, note that 
\be
\Bigl  [ k,P_{1},L_{12},P_{2},L_{23},P_{3}\Bigr ] _{\*} 
=3i\hbar^3 \Bigl ( P_2 \* \{ \! \{k,H_{qm} \}\! \} 
+\{\! \{ k,H_{qm}\}\! \} \* P_2 \Bigr ) + {\cal Q} ~,
\ee
where ${\cal Q}$ is an $O(\hbar^5)$ sum of triple commutators of $k$ with 
invariants. Consequently, the proper quantization of 
(\ref{omunene}) is 
\be
\Bigl  [k,P_{1},L_{12},P_{2},L_{23},P_{3}\Bigr ] _{\*} =
3i\hbar^3  \frac{d}{dt} \Bigl  ( P_2 \* k + k\* P_2\Bigr )+ {\cal Q} ~,
\ee
and again reduces to (\ref{omunene}) in the $\hbar\rightarrow 0$ limit,
as ${\cal Q}$ is subdominant in $\hbar$ to the time derivative term.
The right hand side not being an unadorned derivation on $k$, it does not 
impose a Leibniz rule analogous to (\ref{qLeibnz}) on the left hand side, 
so it fails the mathematical desiderata mentioned, at no compromise to its
validity, however.  The $N>3$ case parallels the above through use of 
fully symmetrized products. 

\section {Conclusions}
The first aim of this paper has been to illustrate the power 
and simplicity of phase-space quantization of superintegrable 
systems which would suffer from operator ordering ambiguities in 
conventional quantization. Many of these $\sigma$-models quantized here,
such as the $S^N$ models, have already been quantized conventionally
\cite{velo,lakshmanan,higgs,leemon}, through elaborate operator algebra
preserving the maximal symmetries of these systems (see especially
the 2nd reference of \cite{winter}). But not all, 
such as the Chiral Models, whose geometrical 
complication has so far only partially yelded to indirect methods 
\cite{chugoddard}.

Here, the procedure of determining the proper
symmetric quantum Hamiltonian has yielded remarkably compact and elegant
expressions, since a survey of all 
alternate operator orderings in a problem with such ambiguities amounts,
in deformation quantization, to a survey of the ``quantum correction" 
$O(\hbar)$ pieces of the respective kernel functions, ie the inverse 
Weyl transforms of those operators, and their study is greatly systematized 
and expedited. Choice-of-ordering problems then reduce to purely
$\*$-product algebraic ones, as the resulting preferred orderings are 
specified through particular deformations in the c-number kernel expressions 
resulting from the particular solution in phase space. For the $N$-spheres,
our results agree with the results 
of \cite{velo,lakshmanan,higgs,leemon,winter}, 
while quantum hamiltonians for chiral models such as (\ref{Plechazunga}) 
are new. With functional methods confined to phase space, we have also 
illustrated how the spectra of such hamiltonians may be obtained.
One might wish to contrast the quantum correction found here in 
(\ref{okanumaihi})
to the free-space angular-momentum quantum correction \cite{dahlP}, 
which is also $O(\hbar^2)$, although a constant, reflecting the vanishing 
curvature of that underlying manifold. Predictably, on the North 
pole of (\ref{okanumaihi}), $u=1$, and these expressions coincide.
This difference and pole coincidence carries over for all dimensions, 
as evident in the quantum correction (\ref{sN}) for $S^N$. 

More elaborate isometries of general manifolds in such models are expected to 
yield to analysis similar to what has been illustrated for the prototypes 
considered here.

The second main conclusion of the paper has 
been a surprising application: 
Quantization of maximally superintegrable systems in phase space has 
facilitated explicit testing of NB 
quantization proposals, through direct comparison to the conventional 
quantum answers thus found. The classical evolution of all 
functions in phase space for such systems is alternatively specified through 
NBs. However, quantization of NBs has been considered problematic 
ever since their inception.
Nevertheless, it was demonstrated that Nambu's early quantization  
prescription \cite{nambu} can, indeed, succeed, despite widespread expectations
to the contrary.  Comparison to the standard Moyal deformation quantization 
utilized in this work vindicates Nambu's early quantization  prescription
(and invalidates other prescriptions), for systems such as $S^N$.
We thus stress the utility of phase space quantization as a comparison testing 
tool for NB quantization proposals. 

\phantom{AA}

\noindent{\Large{\bf Acknowledgments}}\newline 
We gratefully acknowledge helpful discussions with R Sasaki, 
D Fairlie, and Y Nutku. 
This work was supported in part by the US Department of Energy, 
Division of High Energy Physics, Contract W-31-109-ENG-38, and the NSF Award 
0073390. 

\phantom{AA}

\appendix{\Large \bf Appendix}

Equations (\ref{Lie}, \ref{karagoz}) 
are implicit in \cite{bcz} and throughout the literature,
but it may be useful here to provide a direct geometrical proof.

For any Vielbein satisfying $g_{ab}=V_{a}^{j}V_{b}^{j}$ 
and the Maurer-Cartan equation 
\be
\partial_{\lbrack a}V_{b]}^{j}+f_{jmn}V_{a}^{m}V_{b}^{n}=0,
\ee
we have ((3.16) of \cite{bcz}) 
\be
D_{a}V^{bi}=-f_{imn}V_{a}^{m}V^{bn},%
\ee
or 
\be 
\partial_{a}V^{bi}=-\Gamma_{\;ac}^{b}V^{ci}-f_{imn}V_{a}^{m}V^{bn},%
\label{BCZ3.16} %
\ee 
where the Christoffel connection is the usual functional of just the metric
$g_{ab}$. These hold for both groups of Vielbeine, for the same metric and 
hence Christoffel connection; and so, for both 
groups of charges $~^{\pm}V^{aj}p_a$, it follows that   
\be 
\left\{  V^{aj}p_{a},V^{bk}p_{b}\right\}   
&=&p_{a}V^{bk}\left(  \partial_{b}V^{aj}\right)  -p_{b}V^{aj}\left( 
 \partial_{a}V^{bk}\right)  \nonumber\\
&=&  p_{b}V^{ak}\left(  -\Gamma_{\;ac}^{b}V^{cj}-f_{jmn}V_{a}^{m}V^{bn}\right)
-p_{b}V^{aj}\left(  -\Gamma_{\;ac}^{b}V^{ck}-f_{kmn}V_{a}^{m}V^{bn}\right)
\nonumber\\
& =&  -2f_{jkn}V^{bn}p_{b}~.\label{beberuhi}%
\ee 
Actually, for {\em } any two Vielbeine, $V$ and $\widetilde{V}$, 
producing the same metric (and hence Christoffel connection) and 
obeying their own Maurer-Cartan equations, both satisfy equations like
(\ref{BCZ3.16}), and hence algebras like (\ref{beberuhi}). 

The cross PBs, however, need not automatically vanish in the general case, 
\be 
\left\{  V^{aj}p_{a},\widetilde{V}^{bk}p_{b}\right\}   &  = & \left(
\widetilde{V}^{ak}\partial_{a}V^{bj}-V^{aj}\partial_{a}\widetilde{V}%
^{bk}\right)  p_{b} \nonumber\\
&  =& \widetilde{V}^{ak}\left(  -\Gamma_{\;ac}^{b}V^{cj}-f_{jmn}V_{a}^{m}%
V^{bn}\right)  p_{b}-V^{aj}\left(  -\Gamma_{\;ac}^{b}\widetilde{V}%
^{ck}-\widetilde{f}_{kmn}\widetilde{V}_{a}^{m}\widetilde{V}^{bn}\right)p_{b}
\nonumber\\
&  =& \left(  -f_{lmn}V_{a}^{m}V_{c}^{n}\delta^{lj}\widetilde{V}^{ak}%
+\widetilde{f}_{lmn}\widetilde{V}_{d}^{m}\widetilde{V}_{c}^{n}\delta
^{lk}V^{dj}\right) g^{cb}p_{b}\nonumber \\
&  =& \left(  -f_{lmn}V_{a}^{m}V_{c}^{n}V_{d}^{l}V^{dj}\widetilde{V}%
^{ak}+\widetilde{f}_{lmn}\widetilde{V}_{d}^{m}\widetilde{V}_{c}^{n}%
\widetilde{V}_{a}^{l}\widetilde{V}^{ak}V^{dj}\right) g^{cb}p_{b}\nonumber \\
&  =& \left(  -f_{lmn}V_{d}^{l}V_{a}^{m}V_{c}^{n}+\widetilde{f}_{lmn}%
\widetilde{V}_{d}^{m}\widetilde{V}_{a}^{l}\widetilde{V}_{c}^{n}\right)
V^{dj}\widetilde{V}^{ak}g^{cb}p_{b} \nonumber\\
&  =& -\left(  f_{lmn}V_{d}^{l}V_{a}^{m}V_{c}^{n}+\widetilde{f}_{lmn}%
\widetilde{V}_{d}^{l}\widetilde{V}_{a}^{m}\widetilde{V}_{c}^{n}\right)
V^{dj}\widetilde{V}^{ak}g^{cb}p_{b}~. 
\ee 
The terms in the parenthesis of the final line are actually the torsions on 
the respective manifolds ((3.10) of \cite{bcz}), induced by the corresponding 
Vielbeine, up to a normalization,
\be
S_{dac} =f_{lmn}V_{d}^{l}V_{a}^{m}V_{c}^{n}, \qquad 
\widetilde{S}_{dac}
=\widetilde{f}_{lmn}\widetilde{V}_{d}^{l}\widetilde{V}_{a}^{m}%
\widetilde{V}_{c}^{n}.%
\ee
Hence,
\be
\left\{  V^{aj}p_{a},\widetilde{V}^{bk}p_{b}\right\}   = 
-\left(S_{dac}+ \widetilde{S}_{dac} \right)
V^{dj}\widetilde{V}^{ak}g^{cb}p_{b}~. 
\ee

However, for the specific chirally enantiomorphic Vielbeine defined above, 
${V}_{a}^{m} = ~^{(+)}{V}_{a}^{m}$ and 
$\widetilde{V}_{a}^{m}=~^{(-)} {V}_{a}^{m}$,
it further follows from eqn (3.10) of \cite{bcz}  that, in fact,  
\be 
\widetilde{S}_{dac} &  = &  -\frac{1}{2}Tr \left ( \partial_{\lbrack d}  
U^{-1}\,U\partial_{a} U^{-1}\,U\partial_{c]}U^{-1}
\,U \right ) \nonumber  \\
& =&  \frac{1}{2}Tr\left(  U^{-1}\partial_{\lbrack d}U\,U^{-1}\partial_{a}U
\,U^{-1}\partial_{c]}U\right)  =-S_{dac}~,
\ee 
and thus, indeed,  (\ref{karagoz}) holds.

\end{document}